**Functional change in children with cerebral palsy**

*Submitted as an Original Article by*

[a]Derek John Curtis, [b]Pauline Holbrook, [b]Sarah Bew, [b]Lynne Ford, [c]Penny Butler

[a] Department of Neurorehabilitation TBI Unit, Rigshospitalet Neurocenter, University of Copenhagen [b] The Movement Centre, The Robert Jones & Agnes Hunt Hospital, Oswestry, Shropshire, UK [c] Manchester Metropolitan University, Manchester, UK

Corresponding author: Derek John Curtis, Department of Neurorehabilitation TBI Unit, Rigshospitalet Neurocenter (Satellite Department at Hvidovre Hospital), University of Copenhagen, Kettegaard Allé 30, DK-2650 Hvidovre Denmark.

Email address:

Derek John Curtis, PhD – corresponding author – derek.john.curtis@regionh.dk

# Abstract

## Introduction

There is increasing focus on the association between trunk control and functional abilities in children with cerebral palsy (CP). The purpose of this study was to determine the extent of functional change in children with CP who participated in specific trunk and head postural control training combined with physical therapy treatment as usual (TAU).

## Methods

This study included 140 consecutive referrals to a centre specialising in head and trunk postural control (Targeted Training (TT)) between 2009 and 2016. Twenty-five children discontinued therapy due to surgery, health, family issues or poor attendance. The remaining 115 children (46 girls, 69 boys) had a mean age of 6 y 6 mo (SD 2 y 8 mo) with participants from all GMFCS levels.

The intervention was a program of TT and ongoing TAU with a mean duration of 11 months.

Gross Motor Function Measure (GMFM), Pediatric Evaluation of Disability Inventory functional skills, Chailey Levels of Ability and Segmental Assessment of Trunk Control were administered before and after the intervention.

## Results

There were significant improvements in all outcomes. GMFM improvements exceeded those predicted from the published reference curves, especially for the children with more severe cerebral palsy.

## Conclusions

Functional improvement exceeded the expected norm, especially in those children with more severe gross motor function disability. The other outcomes also showed significant improvements. These findings support the case for further studies and, if needed, tool development to facilitate determination of the critical elements in a combined therapy approach of TT with TAU.

1. Introduction

Postural control deficits play a central role in the motor disabilities of children with cerebral palsy (CP)[1] with trunk and head postural control having a significant influence on gross and fine motor function[2–4]. Developmental changes in sitting stability in typically developing (TD) infants have been identified as specific to the region of the trunk being investigated[5] while improved reaching has been demonstrated in children with CP and trunk dysfunction when external assistance is given to match their intrinsic level of trunk control[6]. A recent study by Pierret and co-authors[7] showed that an axial (trunk) rehabilitation programme had a positive impact on seated trunk control in children with mild CP (GMFCS I and II). El Shemy[8] found that the addition of core stability exercises to a treatment program improved both trunk muscle endurance and gait characteristics in children with hemiplegic CP. Yildiz and co-authors reported a significant positive correlation between trunk control and upper extremity functions in 32 children with bilateral CP[9]. This focus on trunk control and its relation to functional activity complements the retrospective cross-sectional study by Curtis and co-authors[10] that suggested a strong association between segmental trunk postural control and gross motor function and mobility in children with CP.

Targeted Training (TT) is one of the training methods that aims specifically at improving postural control of the head and trunk[11,12]. It uses a sequential and segmental approach to axial control development, as proposed by Butler and Major[12] and recommended by Saavedra and Woollacott[13]. Specialised equipment is used to support the child in an optimal vertical position and provides firm support directly beneath the highest segmental level at which control is poor or not demonstrated; this is determined using the Segmental Assessment of Trunk Control (SATCo)[14]. The subsequent training consists of games and activities that challenge static and active control of those free joints/segments above the support level. Progression in training is achieved either by lowering the support level so the child has to exercise postural control over a greater number of segments and/or by increasing the postural control challenge by setting the equipment on a rocking base to introduce instability, thus training reactive control.

Two studies have been published reporting the effect of TT. The most recent study[4] was a randomised controlled trial with 28 children with moderate to severe CP randomised to treatment as usual (TAU) or TT. The study reported no difference between groups in either Pediatric Evaluation of Disability Inventory (PEDI) or Gross Motor Function Measure (GMFM) after a six month TT intervention. An earlier study[15] was a case series of six children with CP aged from 2 years and 5 months to 7 years and 5 months without independent sitting balance who were studied under a random order of three conditions: TT, sham TT and TAU. The attainment of independent sitting balance and the SATCo test were used as the outcome measures. All six children showed an increase in their movement control and attained independent sitting balance within 12-25 weeks. The study concluded that TT may be an effective means of promoting movement control and functional ability.

The lack of published reports on the effect of TT and the conflicting evidence from these two publications illustrate the need for further studies in this area. Controlled experiments clearly have a role in that a given therapy is delivered under controlled conditions, but they are costly to resource adequately to produce quantifiable change, particularly with respect to numbers of participants, dosage, and the length of the intervention period. Thus the first stage in evaluating a specific therapy approach is to demonstrate whether, in a clinical setting, functional change occurs and if that functional change exceeds what might be expected from published data, such as the GMFM reference percentiles[16]. This first stage, in itself, can be difficult since, in a clinical setting, children may concurrently be receiving more than one form of intervention. However, if justified, future work can then elucidate the mechanisms of the various intervention components. The purpose of this study was therefore to analyse clinical data from a specialised referral centre located in the United Kingdom to determine the extent of functional change in children with CP who participated in TT combined with TAU.

## 2. Materials and methods

Participants in the study were all consecutive referrals of children with CP to a single specialist treatment centre between 2009 and 2016. Children were excluded from the study if they were referred and assessed but were very poor or non-attenders or withdrew from treatment due to surgery, health or family issues. If children were referred more than once during this period, then only the first course of TT was included.

The children received a combination of TAU and TT. The TT took place at the child's home or school and was supplemental to the child's normal therapy. Clinical decisions concerning the form and progression of the therapy were made in accordance with the published algorithm[11]. Therapy at the centre is nominally prescribed for a period of nine months but is extended to accommodate ability of the family to attend on specific dates or because of intervening illness/surgery. If a child shows functional improvement, this results in a further course of therapy rather than extension of a course. Training logbooks are not used by the centre, but the parents or carers are asked to train with the child for 30 minutes a day on five or six days each week. Training consists of motivating activities that exercise and challenge the child's postural control at the targeted segment utilising custom-designed positioning equipment. The child's parents were shown and practiced these specific TT activities with their child at the time of equipment supply. The TAU programme was under the direction of each child's local physical therapist and was generally described as 'eclectic'.

GMFM, SATCo, the functional skills dimension of the PEDI and the Chailey Levels of Ability (CLA) were administered at the centre at the start and end of the combined TT/TAU programme by physical therapists with a minimum of two years paediatric experience. The data was analysed for significant changes in scores from baseline to completion of the course TT/TAU.

A supplementary analysis was made of the GMFM data by comparing the outcomes for TT/TAU to GMFM changes calculated using the published tabulated reference percentiles[16]. This was done by determining the percentile for each child at the start of their TT/TAU intervention with reference to their age, GMFCS level and GMFM score. The expected change in GMFM was then calculated using the percentile curve and the training period and this change used as a control value in a supplementary statistical analysis of the effect of TT/TAU. A similar method of controlling for age using the motor growth curves has previously been reported by Yabunaka[17] although in this earlier study the published reference percentiles were not used.

Statistical method
GMFM data were checked for normal distribution prior to statistical analysis using Shapiro Wilkes tests. The ordinal data of the SATCo, PEDI and CLA scores were analysed using non-parametric statistical methods (Wilcoxon signed-rank test). Differences between the clinical changes in GMFM and the changes calculated from the tabulated reference percentiles were not normally distributed, so these data were also analysed using a Wilcoxon signed-rank test.
Statistical analysis was performed using IBM© SPSS© Statistics version 24. A p-value of less than 0.05 was considered statistically significant.

## 3. Results

A total of 140 children (56 girls and 84 boys) were included in the study. Of these 140 children, 25 were excluded from the analysis for the following reasons: surgery (3), health (6), family issues (5), poor attenders (6), non-attenders (5). These excluded children were not statistically different from the children included in the study on any of the group demographics. The analysis therefore included the remaining 115 children (46 girls and 69 boys) (Table 1). The children had a mean age of 6 y 6 mo (SD 2 y 8 mo). Ninety-two of the children had spastic CP, seven had ataxic and 16 dyskinetic. The children trained for a mean period of 11 (SD 2) months. There was a highly significant improvement in GMFM between baseline and final scores following TT/TAU (*Mdn* 3.2, Q1-Q3 1.5-6.8, $p<0.001$). The median change scores for each GMFCS category are shown in table 1.

Change scores for the SATCo showed a significant improvement in trunk control from baseline (*Mdn=learning static, active and reactive control at the Lower Thoracic segment)* to final score (*Mdn= learning static, active and reactive control at the Lower Lumbar segment)*, $p<0.001$ (table 2). Analysis by GMFCS level showed significant improvements in SATCo scores for all GMFCS levels apart from level I.

Change scores for the PEDI showed a significant improvement in functional skills scaled scores from baseline (*Mdn*=33) to final score (*Mdn*=40), $p<0.001$ and caregiver scaled scores from baseline (*Mdn*=34) to final score (*Mdn*=41), $p<0.001$ for all children. Analysis by GMFCS level showed significant improvements in PEDI scores for all GMFCS levels apart from level I (table 3).

There were significant improvements in the children's abilities post training measured using the CLA (Table 4). This was true for all five dimensions of the test. The greatest magnitude of change within the individual GMFCS levels occurred in the sitting and standing dimensions. Change did not reach statistical significance for GMFCS I and the lying dimensions for GMFCS II.

Seventeen children were excluded from the supplementary GMFM analysis as they were either younger than two years when they started their TT/TAU or over 12 when they finished: the age range of the reference percentiles used to generate the hypothetical controls is from two years to 12 years. The 98 children included in this analysis comprised 40 girls and 58 boys. The children had a mean age of 5 y 7 mo (SD 2 y 1 mo). Seventy-nine of the children had spastic CP, six had ataxic and 13 dyskinetic. There were 3, 14, 26, 30 and 28 percent of the children respectively in GMFCS levels I to V.

Table 5 shows results from the analysis showing the difference between real changes and the corresponding change calculated from tabulated reference percentiles in the GMFM scores.

## 4. Discussion

This retrospective study has demonstrated a statistically significant improvement in functional change (GMFM, SATCo, PEDI and CLA) in children with CP who participated in a combined therapy programme of TT with TAU. The median change of GMFM was 3.2 points for the entire study group. A number of the children in the study group are under the age at which their GMFM is expected to plateau or fall and the supplementary analysis was performed as an attempt to control for age: this showed a significant median change of 1.9 GMFM points.

Analysis of the individual GMFCS levels shows significant improvements for all GMFCS levels with the exception of level I. It is possible that this exception is due to a type II error related to the very small sample size (n=3). In the supplementary analysis, changes in GMFCS level II did not reach significance. This is possibly due to the relatively large increase predicted by the tabulated reference percentiles for GMFCS level II compared with the higher GMFCS levels. This change from significance to non-significance in GMFCS II illustrates the need for caution in interpreting the results of interventions in young children and infants and children in the lower GMFCS levels where the intervention is lengthy and there is no control group.

It is interesting to note that the largest significant increase in median GMFM is for children in GMFCS V. In the initial analysis, this median increase is 3.4 points and when controlled in the supplementary analysis the increase is 3.0 points. This value is three times the size of the expected change using the tabulated reference values and is significant. It is also a considerable increase from the baseline median GMFM of 21.3 points for GMFCS V.

Significant changes in the other GMFCS levels are around a median of 3 points, but when adjusted using the reference percentiles, significant improvements are only seen in GMFCS III-IV and are 1.5 and 0.9 points respectively. In the light of these results, it would appear that TT/TAU has greatest effect on gross motor function for children in GMFCS III-V, but that the largest effect both absolute and relative to the initial GMFM score is seen in GMFCS level V.

The minimum clinically important difference (MCID) for the GMFM has not been published for GMFCS IV-V, but a single study[18] has determined MCID for ambulatory children. This study reported an MCID of 0.8 GMFM points for an effect size of 0.5 and 1.3 GMFM points for an effect size of 0.8 based on a sample of 292 children in GMFCS I-III. The reported MCID in GMFM score decreased with higher GMFCS level; it is therefore possible that a smaller change in GMFM than 0.8 points would produce a clinically important difference for children in GMFCS IV and V.

The statistical increase in trunk control, as measured by SATCo, showed a median increase of two trunk segments for active and reactive control and one segment for static control: this provides further confirmation of the relation between trunk control and functional abilities in children with CP, as previously reported.[7-10] Analysis of the individual GMFCS levels shows, as with GMFM, that there are significant improvements for all GMFCS levels with the exception of level I. Improvements are in the order of two segmental levels for GMFCS II-IV and one segmental level for GMFCS V.

PEDI functional skills dimension and caregiver assistance scaled scores increased significantly by a median of 4 points and 1 point respectively. Analysis of the change showed a similar pattern to GMFM with a significant increase in PEDI score for all GMFCS levels with the exception of GMFCS level I. The median score changes for functional skills were similar for all GMFCS levels with the exception of GMFCS level V. There was a greater diversity in changes in median functional caregivers scores but, as with functional skills, no change in the median score for GMFCS V. This could be explained by the lack of a direct link between the gross motor functional improvements seen in these children and the transfer of these improved functions to activities of daily living assessed within the PEDI. There are unfortunately no published MCID studies for the PEDI test when administered on children with CP, so it is not possible to assess the importance of the changes in PEDI scores that are presented in this study.

There are significant changes in CLA but the median changes are zero for all GMFCS levels in supine and prone lying. Median changes of one level occurred in floor sitting for GMFCS II-IV, box sitting for GMFCS II-III and standing for GMFCS III-IV. It is likely that GMFCS I experience a ceiling effect with this test. GMFCS II-IV appear to improve in their levels of ability when they have an upright trunk (sitting and standing). This is possibly related to the focus of the TT intervention with its direction towards improving control of the upright posture[12,15]: this focus may also have been part of the TAU. The lack of change in children in GMFCS V reflects once again the challenges in achieving such substantial improvements in postural control in this group of children in the short to medium term.

It appears from the results of this study that the best effect of TT/TAU is in GMFCS levels III-V. A review of exercise interventions to improve postural control in children with CP[19] identified 13 interventions from 45 studies. Five of these interventions were supported by a moderate level of evidence: gross motor task training, hippotherapy, treadmill training with no body weight support (no-BWS), trunk-targeted training, and reactive balance training. This systematic review identified three studies reporting outcomes for children in GMFCS IV-V. These studies reported the effect of two interventions: hippotherapy[20] and training with hippotherapy simulators[21,22]. Hamill et al[20] reported no change in the level of sitting scale or GMFM-88 score for three children in GMFCS level V following 10 weekly sessions of 50 minutes hippotherapy. Silva e Borges et al.[21] reported a reduced sway in sitting with no change in GMFCS [sic] in a group of 40 children in GMFCS II-IV. Herrero et al.[22] reported no change in the sitting assessment scale or the GMFM-66 score but an improvement in GMFM: B (sitting) in a group of 38 children with CP GMFCS I-IV. In contrast, Martín-Valero and co-authors[23] reported gains in postural alignment and head and trunk balance in their narrative review of hippotherapy in children with CP. The focus on the head and trunk postural control of hippotherapy may have contributed to these gains although they are not conclusive.

The present study has a number of limitations. There is an inherent risk of bias as the data are collected at a treatment centre that specialises in TT. The recommendations for TT are five or six sessions of 30 minutes per week, however we cannot be certain that this was the training intensity for the participants in this study as the training took place at home or in the children's school or kindergarten. The relationship between dose and response for this training method is unknown, so departures from the recommended training dose could have an effect on the outcome. The detail of the TAU programme is also unknown, as is the adherence to recommended dose. The method of controlling for development of the children using the Canadian cohort assumes that the participants in this study could expect to develop in a similar way. There may be cultural and resource differences between the two countries that affect the motor development of the children so the Canadian cohort do not reflect the motor development of children with CP in the United Kingdom. However, the purpose of this study was to identify change, rather than attribute that change to specific therapy components. There is also the possibility of mass significance when using so many statistical procedures, although the levels of significance would seem to indicate that the chance of a type I error is small. The evidence for GMFCS I is very inconclusive as there are only three children at this level.

This study also has some strengths. One of the principal strengths is that the results reflect the clinical reality of TT/TAU as a treatment form and not the measured effect in a research setting. Although the data are collected in a clinical setting, they have been collected systematically by a small group of experienced paediatric physical therapists. This study attempts to control for time as a confounder by using data from the CanChild cohort and despite this correction there still seems to be a clinically meaningful significant effect of the intervention.

## 5. Conclusions

This study systematically documents positive clinical outcomes in a group of children with CP who have undergone a course of TT/TAU. The study was not set up to determine the cause of change or to attribute that change to any specific therapy approach, but simply to examine the extent of functional change. It was found that, with a combination of Targeted Training and treatment as usual, children with CP experienced a greater change in function than the expected norm (3.2 points on GMFM) and that this improvement was especially large in those children with greater gross motor function disability. The precise cause of the change in function remains open. However, these encouraging results provide motivation and justification to develop diagnostic tools for accurate and objective measurement of changes in trunk control, in a form that is standardised across clinicians, across centres and across differing therapeutic approaches. This would both facilitate determination of the critical elements in this combined therapy approach of Targeted Training with treatment as usual and enhance understanding of therapy mechanisms and outcomes.

Table 1 Change in GMFM scores by GMFCS level

| GMFCS | n | Baseline | Post TT | Median value of Change | Sig. |
|---|---|---|---|---|---|
| All | 115 | 37.8 (26.0-49.2) | 45.9 (30.6-45.9) | 3.2 (1.5-6.8) | <0.001** |
| I | 3 | 74.2 (61.2-74.8) | 74.8 (69.6-86.5) | 8.4 (0-12.4) | 0.180 |
| II | 15 | 58.8 (48.5-65.0) | 62.7 (51.6-70.0) | 3.2 (1.5-5.8) | 0.003** |
| III | 27 | 47.7 (45.6-50.6) | 50.6 (48.5-53.6) | 2.8 (1.7-5.4) | <0.001** |
| IV | 37 | 36.4 (28.0-46.0) | 41.4 (33.9-49.2) | 3.1 (1.0-7.4) | <0.001** |
| V | 33 | 21.3 (19.7-26.0) | 26.0 (22.7-30.9) | 3.4 (1.4-7.6) | <0.001** |

* $p<0.05$ ** $p<0.01$. Values are median (Q1-Q3)

Table 2 Change in SATCo static, active and reactive scores by GMFCS level

| GMFCS | n | Baseline | Post Therapy | Median value of Change | Sig. |
|---|---|---|---|---|---|
| SATCo static | | | | | |
| All | 115 | 4 (2-6) | 6 (5-8) | 1 (0-2) | <0.001** |
| I | 3 | 5 (3-8) | 6 (6-8) | 1 | 0.180 |
| II | 15 | 7 (4-8) | 8 (7-8) | 1 (1-2) | 0.007** |
| III | 27 | 5 (4-7) | 7 (6-8) | 1 (0-2) | <0.001** |
| IV | 37 | 4 (3-6) | 6 (5-8) | 2 (1-2.5) | <0.001** |
| V | 33 | 2 (1-2.5) | 3 (1.5-5) | 1 (0-2) | <0.001** |
| SATCo active | | | | | |
| All | 115 | 4 (2-6) | 6 (4-8) | 2 (0-2) | <0.001** |
| I | 3 | 5 (3-8) | 6 (6-8) | 1 | 0.180 |
| II | 15 | 6 (4-8) | 8 (7-8) | 1 (0-3) | 0.003** |
| III | 27 | 5 (4-7) | 7 (6-8) | 1 (0-2) | <0.001** |
| IV | 37 | 4 (3-6) | 6 (5-8) | 2 (1-3) | <0.001** |
| V | 33 | 1 (1-2.5) | 3 (1-4.5) | 1 (0-2) | <0.001** |
| SATCo reactive | | | | | |
| All | 115 | 4 (3-6) | 6 (4-8) | 2 (1-2.5) | <0.001** |
| I | 3 | 5 (3-8) | 6 (6-8) | 1 | 0.180 |
| II | 15 | 6 (4-7) | 8 (6-8) | 2 (0-3) | 0.003** |
| III | 27 | 5 (4-6) | 7 (6-8) | 1 (0-2) | <0.001** |
| IV | 37 | 4 (3-6) | 6 (5-8) | 2 (1-2.5) | <0.001** |
| V | 33 | 2 (1-3) | 3 (1.5-5) | 1 (0-3) | <0.001** |

* $p<0.05$ ** $p<0.01$. Values are median (Q1-Q3) The numbers represent median values of the SATCo trunk segmental level at which control was being learnt: 1= head control, 2= upper thoracic level, 3= mid-thoracic, 4= lower thoracic, 5= upper lumbar, 6= lower lumber, 7= full trunk control, and 8= full trunk control achieved. Note that the non-integral numbers reported were purely for statistical purposes. In real life situations, no half-level would be credited.

Table 3 Change in PEDI functional skills and functional skills caregiver scaled scores by GMFCS level

| GMFCS | n | Baseline | Post TT | Median value of Change | Sig. |
|---|---|---|---|---|---|
| PEDI functional skills | | | | | |
| All | 115 | 33 (21-50) | 40 (23-56) | 4 (0-8) | <0.001** |
| I | 3 | 67 (44-67) | 67 (50-73) | 6() | 0.157 |
| II | 15 | 55 (46-63) | 59 (49-72) | 5 (3-14) | 0.001** |
| III | 27 | 42 (49-53) | 49 (53-59) | 4 (2-9) | <0.001** |
| IV | 37 | 32 (25-42) | 31 (37-48) | 5 (1-7) | <0.001** |
| V | 33 | 18 (15-22) | 15 (23-27) | 0 (0-7) | <0.001** |
| PEDI functional skills caregiver | | | | | |
| All | 115 | 34 (0-51) | 41 (12-55) | 1 (0-9) | <0.001** |
| I | 3 | 57 (47-62) | 57 (51-70) | 4 () | 0.180 |
| II | 15 | 57 (44-65) | 57 (50-70) | 2 (0-7) | 0.005* |
| III | 27 | 39 (47-55) | 47 (54-59) | 5 (0-12) | <0.001** |
| IV | 37 | 32 (12-42.5) | 20 (37-52) | 2 (0-10) | 0.001** |
| V | 33 | 0 (0-12) | 0 (0-20) | 0 (0-6) | 0.028* |

* *p*<0.05 ***p*<0.01. Values are median (Q1-Q3)

Table 4 Change in Chailey Levels of Ability for all children by GMFCS level

| GMFCS level | n | Supine lying | | Prone lying | | Floor sitting | | Box sitting | | Standing | |
|---|---|---|---|---|---|---|---|---|---|---|---|
| | | Median Change | Sig | Median Change | Sig | Median Change | Sig | Median Change | Sig | Median Change | Sig |
| All | 115 | 0 (0-1) | <0.001** | 0 (0-1) | <0.001** | 0 (0-1) | <0.001** | 0 (0-1) | <0.001** | 0 (0-1) | <0.001* |
| I | 3 | 0 (0-0) | 1.000 | 0 (0-0) | 1.000 | 0 (0-0) | 1.000 | 0 (0-0) | 1.000 | 0 (0-0) | 1.000 |
| II | 15 | 0 (0-0) | 0.102 | 0 (0-0) | 0.109 | 1 (0-4) | 0.011** | 1 (0-2) | 0.011* | 0 (0-4) | 0.027* |
| III | 27 | 0 (0-0) | 0.038* | 0 (0-1) | 0.016* | 1 (0-1) | 0.001** | 1 (0-4) | <0.001** | 1 (0-3) | <0.001** |
| IV | 37 | 0 (0-1) | 0.002** | 0 (0-1) | <0.001** | 1 (0-1) | <0.001** | 0 (0-1) | <0.001** | 1 (0-1) | <0.001** |
| V | 33 | 0 (0-1) | 0.001** | 0 (0-1) | 0.001** | 0 (0-1) | 0.002* | 0 (0-1) | 0.005** | 0 (0-0) | 0.025* |

\* *p*<0.05 \*\**p*<0.01. Values given are median (Q1-Q3) for change scores. Positive change scores represent an increase in level from baseline to post training.

Table 5 Real change and change from tabulated reference percentiles in GMFM for the children in the supplementary analysis

| GMFCS level | n | Real change | Median Change from tabulated reference percentiles | Difference | Sig. |
|---|---|---|---|---|---|
| All | 98 | 2.8 (1.4-5.6) | 1.5 (0.6-2.5) | 1.6 (-0.1-3.7) | <0.001** |
| I | 3 | 8.4 (0-12.4) | 4.1 (-0.4-5.7) | 2.7 (0.35-8.2) | 0.109 |
| II | 14 | 3.0 (1.4-5.6) | 3.0 (2.7-3.8) | -0.3 (-1.8-2.9) | 0.826 |
| III | 25 | 2.7 (1.6-5.1) | 1.3 (0.3-1.9) | 1.5 (0.5-3.7) | <0.001** |
| IV | 29 | 2.5 (0.9-4.8) | 1.9 (0.2-3.0) | 0.9 (-0.3-3.7) | 0.023* |
| V | 27 | 3.4 (1.4-7.6) | 1.0 (0.6-1.6) | 3.0 (1.2-6.9) | <0.001** |

* $p<0.05$ ** $p<0.001$. Values are median (Q1-Q3)